\documentclass[aps,prl,twocolumn,showpacs,groupedaddress]{revtex4}

\usepackage{graphicx}

\def\Datm{\Delta m_{atm}^2}
\def\Dsol{\Delta m_{sol}^2}
\def\Dcho{\Delta m_{chooz}^2}
\def\mns#1{m_{\nu#1}^2}
\def\sinatm{\sin^22\theta_{atm}}  
\def\sinsol{\sin^22\theta_{sol}}  
\def\sincho{\sin^22\theta_{chooz}}  
\def\e3s{\eta^2_{e3}}

\def\PRC#1#2#3{Phys. Rev. {\bf C#1}, #2 (#3)}
\def\PRD#1#2#3{Phys. Rev. {\bf D#1}, #2 (#3)}
\def\NPB#1#2#3{Nucl. Phys. {\bf B#1}, #2 (#3)}

\def\PLB#1#2#3{Phys. Lett. {\bf B#1}, #2 (#3)}
\def\PRL#1#2#3{Phys. Rev. Lett. {\bf #1}, #2 (#3)}

\def\AJ#1#2#3{Astrophys. J. {\bf #1}, #2 (#3)}

\begin{document}

\preprint{
}

\title{
Antisymmetric Higgs representation in SO(10) for neutrinos
}


\author{
Noriyuki Oshimo
}
\affiliation{
Institute of Humanities and Sciences {\rm and} Department of Physics \\
Ochanomizu University, Tokyo, 112-8610, Japan
}


\date{\today}

\begin{abstract}
     A Model based on SO(10) grand unified theory (GUT) and supersymmetry 
is presented to describe observed phenomena for neutrinos.  
The large mixing angles among different generations, 
together with the small masses, are attributed to the Higgs boson 
structure at the GUT energy scale.   
Quantitative discussions for these observables are given, taking 
into account their energy evolution.   
\end{abstract}


\maketitle


     Experiments for atmospheric \cite{atm} and solar \cite{sol} neutrinos 
suggest non-vanishing but extremely small masses for the neutrinos.
Also observed are large mixing angles among different generations for 
the leptons.  
These results imply existence of physics beyond the standard model.  
In fact, the lightness of the neutrinos could be naturally accounted for
by those models based on grand unified theory (GUT) which contain
large Majorana masses for the right-handed neutrinos.

     The GUT models, however, face one serious problem posed by the large 
generation-mixing angles of the leptons.  
Since the leptons stand on an equal footing with the quarks, in simple models 
the amount of their generation mixings becomes similar to that 
of the quarks, leading to small mixing angles.  
Some complication is thus necessary for accommodating the observed 
properties of the neutrinos within the framework of GUT.  
Aiming at construction of a plausible model, 
various works have been performed \cite{dorsner}.  

     In this letter we propose an SO(10) GUT model coupled to 
supersymmetry, in which the lepton mixings and the neutrino masses 
can be described fairly simply.  
For group representations which contain the Higgs bosons responsible for 
the quark and lepton masses, our model includes $\bf 120$, 
as well as $\bf 10$ and $\overline {\bf 126}$.  
This $\bf 120$ representation makes the mixing structures different  
between the quarks and the leptons, while the $\overline {\bf 126}$
representation yields large Majorana masses for 
the right-handed neutrinos.  
Below the GUT energy scale, the model is the same as the minimal 
supersymmetric standard model (MSSM) except that    
the dimension-five operators are generated for the SU(2) doublets of 
the leptons and the Higgs bosons.  
These terms give small Majorana masses to the left-handed neutrinos 
after electroweak symmetry is broken down.  
We make quantitative analyses for the neutrino properties at the electroweak 
energy scale, taking into account the energy evolution 
of the mass and mixing parameters for the quarks and leptons.  

     Experiments observe a solar neutrino deficit and an atmospheric 
neutrino anomaly, which could be understood as neutrino oscillations.     
For explaining the atmospheric neutrino problem, the mass-squared difference 
and the mixing angle between the $\mu$-neutrino and the other 
neutrino should be given by  
\begin{eqnarray}
 \Datm &=& (1-8)\times10^{-3} {\rm eV^2},  \nonumber \\ 
 \sinatm &>& 0.85.  
\label{expatm}
\end{eqnarray}
The solar neutrino problem is explained by several scenarios, 
among which most favored by experiments is a large mixing angle 
between the $e$-neutrino and the other neutrino 
under the Mikheyev-Smirnov-Wolfenstein (MSW) effect.  
Assuming this scenario, their mass-squared difference and mixing angle  
are observed as 
\begin{eqnarray}
 \Dsol &=& (2-10)\times10^{-5} {\rm eV^2},  \nonumber \\  
 \sinsol &=& 0.5-0.9.  
\label{expsol}
\end{eqnarray}
These two experimental results suggest that the leptons of three 
generations are fully mixed, contrary to the quark mixings.   
On the other hand, the reactor experiment by CHOOZ \cite{chooz} measures  
the mixing angle between the $e$-neutrino and the other neutrino.  
Non-observation of the mixing excludes the region  
\begin{eqnarray}
 \Dcho &>& 1\times 10^{-3} {\rm eV^2},  \nonumber \\
 \sincho &>& 0.2.   
\label{expchooz}
\end{eqnarray}
The three experimental results provide a guide to constructing a model.  

     The minimal group for GUT which has a representation
for the right-handed neutrinos is SO(10).  
The quark and lepton superfields of one generation, 
both left-handed and right-handed components, are all contained 
in one spinor $\bf 16$ representation.  
The representations which can couple to ${\bf 16}\times{\bf 16}$ 
are $\bf 10$, $\overline{\bf 126}$, and $\bf 120$.  
The Higgs superfields for the masses of the quarks and 
leptons must be in these representations.   

     In the possible three representations for the Higgs superfields, 
the right-handed neutrinos can receive Majorana masses  
only from $\overline{\bf 126}$, provided that a non-vanishing vacuum 
expectation value (VEV) is generated for the scalar field of 
the SU(5) singlet component.   
If SO(10) is broken down to SU(5) by this VEV, its magnitude is as large 
as the GUT energy scale.  
Although the $\overline{\bf 126}$ representation also has SU(2) 
doublet components which could give Dirac masses to 
the quarks and leptons, the large mixing angles for the leptons 
are not yielded by themselves.  
Even if the $\bf 10$ representation is introduced for the 
Higgs bosons, it is difficult to consistently accommodate the 
three experimental results \cite{babu}.  
A scenario with a small mixing angle in the MSW effect for 
the solar neutrino 
oscillation could be provided by incorporating two superfields of $\bf 10$ 
and the Majorana masses for the left-handed neutrinos from the VEV 
of the SU(2) triplet Higgs boson \cite{brahmachari}.  
 
     Large mixing angles for the leptons are not easily obtained 
by $\bf 10$ and $\overline{\bf 126}$.  
This is because every SU(2) doublet Higgs boson in these 
representations gives the same contribution to the quark mixings and 
to the lepton mixings.  
On the other hand, the $\bf 120$ representation has four SU(2) doublet 
components, in which one doublet gives Dirac masses only to 
the neutrinos and another only to the up-type quarks.  
These Higgs bosons could become an origin of the difference 
between the quarks and the leptons.  
 
     We introduce one superfield for each of $\bf 10$, $\bf 120$, 
and $\overline{\bf 126}$.  
All the possible Higgs couplings for the quark and lepton superfields 
at the GUT energy scale are written, 
in the framework of SU(3)$\times$SU(2)$\times$U(1), as 
\begin{eqnarray}
& & \eta^{ij}\left[\phi^{\bar 5}_{10}\left(Q^iD^{cj}+L^iE^{cj}\right)
             +\phi^5_{10}\left(Q^iU^{cj}+L^iN^{cj}\right)\right] \nonumber \\
&+& \epsilon^{ij}\left[\phi^{\bar 5}_{120}\left(Q^iD^{cj}+L^iE^{cj}\right)
                        +\phi^5_{120}L^iN^{cj} \right. \nonumber \\
 & & \left. +\phi^{\overline{45}}_{120}\left(Q^iD^{cj}-3L^iE^{cj}\right)
                        +\phi^{45}_{120}Q^iU^{cj}\right] \nonumber \\
&+& \zeta^{ij}\left[\phi^{\overline{45}}_{\overline{126}}
       \left(Q^iD^{cj}-3L^iE^{cj}\right) 
   +\phi^{15}_{\overline{126}}L^iL^j \right.  \nonumber \\
 & & \left. +\phi^5_{\overline{126}}\left(Q^iU^{cj}-3L^iN^{cj}\right) 
                      +\phi^1_{\overline{126}}N^{ci}N^{cj} \right].
\label{couplings}
\end{eqnarray}
Here, $\phi$'s stand for Higgs superfields with upper and lower indices 
showing transformation properties under SU(5) and SO(10), respectively.
Superfields for the quarks and leptons are denoted in a self-explanatory 
notation by $Q^i$, $U^{ci}$, $D^{ci}$, $L^i$, $E^{ci}$, and $N^{ci}$, 
where the index $i$ represents the generation.  
The group indices are understood.  
The coupling constants $\eta^{ij}$ and $\zeta^{ij}$ are symmetric for the
generation indices, while $\epsilon^{ij}$ are antisymmetric.

     The scalar component of $\phi^1_{\overline{126}}$ is assumed to
have a VEV of order of a GUT energy scale.  
Large Majorana masses are then induced for the right-handed neutrinos 
through $\phi^1_{\overline{126}}N^{ci}N^{cj}$.  
Exchanging the right-handed neutrinos, the couplings 
$\phi^5_{10}L^iN^{cj}$, $\phi^5_{120}L^iN^{cj}$, and 
$\phi^5_{\overline {126}}L^iN^{cj}$ 
lead to dimension-five operators which are composed of 
SU(2) doublet fields for the left-handed leptons and Higgs bosons.  
At the electroweak energy scale where the SU(2) doublet Higgs bosons 
have non-vanishing VEVs, these dimension-five operators become tiny 
Majorana mass terms for the left-handed neutrinos.  
The left-handed neutrinos could also receive Majorana masses from 
$\phi^{15}_{\overline{126}}L^iL^j$.  
However, the VEV of $\phi^{15}_{\overline{126}}$ has to be as small as 
the observed neutrino masses, which necessitates an extreme 
fine-tuning of the Higgs potential.  
We therefore take $\phi^{15}_{\overline{126}}$ for  
enough heavy not to develop a non-vanishing VEV.  
 
     The SU(2) doublet Higgs superfields for 
electroweak symmetry breaking are given by linear combinations 
of the superfields with the same quantum numbers in $\bf 10$, 
$\bf 120$, $\overline{\bf 126}$, or other representations.  
The MSSM Higgs superfields $H_1$ and $H_2$ 
with hypercharges 1/2 and $-1/2$, respectively, are expressed by
\begin{eqnarray}
H_1 &=& C_1^1\phi^{\bar 5}_{10}+C_1^2\phi^{\bar 5}_{120}
          +C_1^3\phi^{\overline{45}}_{120}
          +C_1^4\phi^{\overline{45}}_{\overline{126}}+...   \\
H_2 &=& C_2^1\phi^5_{10}+C_2^2\phi^5_{120}+C_2^3\phi^{45}_{120}
                 +C_2^4\phi^5_{\overline{126}}+...,
\end{eqnarray}
where dots stand for possible components  
belonging to the representations different from $\bf 10$, $\bf 120$, 
and $\overline{\bf 126}$.  
For instance, one superfield of $\bf 126$ is included 
to keep the VEV of the SO(10) $D$ term small by cancellation between 
the VEVs for $\phi^1_{\overline{126}}$ and  $\phi^1_{126}$.  
This $\bf 126$ representation contains SU(2) doublets which become 
components of $H_1$ or $H_2$.       
The gauge coupling unification of SU(3)$\times$SU(2)$\times$U(1) suggests 
that there should exist only one pair of light Higgs doublets.  
We assume that the other linear combinations of the SU(2) doublets 
have large masses and decouple from theory below the GUT energy scale.

     The superpotential of our model relevant to the quark and lepton 
masses are given by  
\begin{eqnarray}
W &=& \eta_d^{ij} H_1Q^iD^{cj} + \eta_u^{ij}H_2Q^iU^{cj}
 + \eta_e^{ij} H_1L^iE^{cj} \nonumber \\
 &+& \eta_\nu^{ij} H_2L^iN^{cj} 
                  + \zeta^{ij}\phi^1_{\overline{126}}N^{ci}N^{cj}.   
\label{superpotential}
\end{eqnarray}
Below the GUT energy scale, the right-handed neutrino superfields $N^{ci}$ 
decouple from theory, owing to their large masses.      
Instead of the last two terms in Eq. (\ref{superpotential}),  
the dimension-five operators are taken into consideration:       
\begin{equation}
 {\cal L} = -\frac{1}{2}\kappa^{ij}
\phi_{H_2}\overline{\psi_{L^i}^c}\phi_{H_2}\psi_{L^j}   + {\rm h.c.},  
\end{equation}
where $\phi_{H_2}$ and $\psi_{L^i}$ represent the scalar component 
of $H_2$ and the fermion component of $L^i$, respectively.   
For definiteness, we define 
the Cabibbo-Kobayashi-Maskawa (CKM) matrix for the quarks and 
the Maki-Nakagawa-Sakata (MNS) matrix for the leptons as  
\begin{eqnarray}
V_{CKM} &=& U_L^{u\dagger}U_L^d,  \\   
 & & U_L^{uT}\eta_uU_R^{u*}=\eta^D_u, \quad 
 U_L^{dT}\eta_dU_R^{d*}=\eta^D_d, \nonumber \\
V_{MNS} &=& U_L^{\nu\dagger}U_L^e, \\   
 & & U_L^{\nu T}\kappa U_L^{\nu}=\kappa^D, \quad 
 U_L^{eT}\eta_eU_R^{e*}=\eta^D_e, \nonumber 
\end{eqnarray}
where $\eta_d^D$, $\eta_u^D$, $\eta_e^D$, and $\kappa^D$ 
denote diagonalized matrices, $U_L$'s and $U_R$'s being unitary matrices.  
Taking the VEVs of electroweak symmetry breaking for positive, 
the diagonal elements of $\eta_d^D$, $\eta_e^D$, and $\kappa^D$ 
should be positive, while those of $\eta_u^D$ should be negative.   
The quarks and leptons then have positive masses.  

     A $3\times 3$ unitary matrix has nine independent parameters.  
Although the numbers of physical parameters in the CKM matrix and 
the MNS matrix are respectively four and six for electroweak interactions, 
more parameters become physical for SO(10) interactions.  
For the expression of $V_{CKM}$ or $V_{MNS}$, we adopt 
the parametrization in which the energy evolution of the independent 
parameters can be traced explicitly \cite{naculich}:   
\begin{eqnarray}
       V &=& P_A^\dagger V_0P_B  \\
       V_0 &=& \left(
        \begin{array}{ccc}
 s_1s_2c_3+{\rm e}^{i\delta}c_1c_2 & c_1s_2c_3-{\rm e}^{i\delta}s_1c_2 & 
                                 s_2s_3  \\
 s_1c_2c_3-{\rm e}^{i\delta}c_1s_2 & c_1c_2c_3+{\rm e}^{i\delta}s_1s_2 & 
                                 c_2s_3  \\
 -s_1s_3 & -c_1s_3 & c_3
        \end{array}
        \right),  \nonumber \\
 && P_A={\rm diag}(\exp(i\sigma_A^1),\exp(i\sigma_A^2),\exp(i\sigma_A^3)), 
                                           \nonumber \\
 && P_B={\rm diag}(\exp(i\sigma_B^1),\exp(i\sigma_B^2),\exp(i\sigma_B^3)), 
                                           \nonumber 
\label{unitary_matrix}
\end{eqnarray}
where $c_i=\cos\theta_i$ and $s_i=\sin\theta_i$.  
For the six arguments $\sigma_A^i$ and $\sigma_B^i$,  
five relative differences are independent.   
Without loss of generality, the angles $\theta_1$, $\theta_2$, 
and $\theta_3$ can be taken to lie in the first quadrant.  

     A large difference between the quark mixings and the lepton 
mixings is expected to occur by the contributions of 
$\phi_{120}^5$ and $\phi_{120}^{45}$.  
For simplicity, we neglect the $\phi_{120}^{\overline{5}}$ and 
$\phi_{120}^{\overline{45}}$ components in $H_1$ and $H_2$.    
Then, at the GUT energy scale, the coefficient matrices for the leptons 
are expressed in terms of those for the quarks as   
\begin{eqnarray}
\eta_e &=& -\frac{3a+b}{a-b}\eta_d+\frac{2}{a-b}(\eta_u+\eta_u^T), 
\label{electrons} \\
\kappa &=& -\eta_\nu\left(M_{\nu_R}\right)^{-1}\eta_\nu^T, 
\label{neutrinos} \\
\eta_\nu &=& -\frac{4ab}{a-b}\eta_d  
     + \frac{a+3b}{2(a-b)}(\eta_u+\eta_u^T)+C_2^2\epsilon, \nonumber \\
M_{\nu_R} &=& \frac{v_R}{C_1^4}
\left[\frac{2a}{a-b}\eta_d -\frac{1}{a-b}(\eta_u+\eta_u^T)\right],  
\nonumber \\ 
a &=& \frac{C_2^1}{C_1^1}, \quad b=\frac{C_2^4}{C_1^4}.  
\nonumber 
\end{eqnarray}
Here, $v_R$ stands for the VEV of $\phi^1_{\overline{126}}$ which 
is of order of the GUT energy scale.  
The coefficients $\eta_u$ and $\eta_d$ are related to each other through  
the CKM matrix, 
\begin{eqnarray}
\eta_d &=& \eta^D_d,  \nonumber \\
\eta_u &=&V_{CKM}^T\eta^D_uUV_{CKM}, \quad  U=U_R^{uT}U_L^u,  
\label{quarks}
\end{eqnarray}
where we have taken a generation basis in which the 
coefficient matrix for the down-type quarks is diagonal.    

     The effect of introducing $\bf 120$ 
is clearly seen if the contribution of $\phi_{120}^{45}$ is neglected.   
Then the matrix $\eta_u$ is symmetric.  
Its diagonalization is made by one unitary matrix ($U_R^u=U_L^{u*})$ and the 
matrix $U$ in Eq. (\ref{quarks}) becomes a unit matrix.  
The coefficient matrices $\eta_d$, $\eta_u$, and $\eta_e$ are 
roughly diagonal simultaneously.  
On the other hand, the matrix $\epsilon$ in Eq. (\ref{neutrinos}) 
has only off-diagonal elements, owing to its antisymmetric property.  
Therefore, the off-diagonal elements for $\kappa$ could be large, 
which would enhance generation mixings for the leptons.    

     The values of $\eta^D_u$, $\eta^D_d$, $\eta^D_e$, $\kappa^D$, 
$V_{CKM}$, and $V_{MNS}$ evolve depending on the energy scale.  
The renormalization group equations for these parameters  
and the gauge coupling constants of SU(3)$\times$SU(2)$\times$U(1) 
close on themselves at the one-loop level.  
Making use of the large mass differences among generations for the quarks 
and the charged leptons, the evolution equations for the independent 
parameters are obtained explicitly \cite{oshimo}.  
Experimentally, the eigenvalues of $\eta_u$, $\eta_d$, 
and $\eta_e$ are known, if the ratio $\tan\beta$ 
of the vacuum expectation values for $H_1$ and $H_2$ is given.  
The CKM matrix elements have also been measured.   
Assuming $C_1^2=C_1^3=C_2^3=0$, therefore, 
unknown quantities at the GUT energy scale are $v_R/C_1^4$, $a$, $b$, 
$C_2^2\epsilon$, and the phase matrices $P_u$ and $P_d$ for $V_{CKM}$.  
If these quantities are given, the values of $\kappa^D$ and 
the MNS matrix are determined at the GUT energy scale and thus 
at the electroweak energy scale.  

\begin{table}
\caption{
The masses of the quarks and leptons (in unit of GeV) and 
the CKM matrix parameters at the electroweak energy scale. 
\label{tab:massmix}
}
\begin{ruledtabular}
\begin{tabular}{cccc}
$m_t$ & $m_c$ & $m_u$ &  \\ 
\hline
 1.8$\times 10^{2}$  & 6.9$\times 10^{-1}$ & 1.8$\times 10^{-3}$ & \\ 
\hline
$m_b$ & $m_s$ & $m_d$ &  \\ 
\hline
 2.9 & 8.5$\times 10^{-2}$ & 1.8$\times 10^{-3}$ & \\ 
\hline
$m_\tau$ & $m_\mu$ & $m_e$ & \\ 
\hline
 1.7 & 1.0$\times 10^{-1}$ & 7.7$\times 10^{-4}$ & \\ 
\hline
$\theta_1$ & $\theta_2$ & $\theta_3$ & $\delta$  \\
\hline
 3.2$\times 10^{-1}$ & 9.0$\times 10^{-2}$ & 4.0$\times 10^{-2}$ & 
 3.0$\times 10^{-2}$  \\
\end{tabular}
\end{ruledtabular}
\end{table}

     Our model is discussed quantitatively.  
We make an assumption that the mass differences among the neutrinos 
are very large, similarly to the quarks or charged leptons.  
Then, the measured quantities by the CHOOZ experiment 
are understood as      
\begin{eqnarray}
    \sincho &=& 4|V_{31}|^2(1-|V_{31}|^2),  \nonumber \\
    \Dcho   &=& \mns3-\mns1,   
\end{eqnarray}
where $m_{\nu i}$ represents the mass eigenvalue for the neutrino 
of the $i$-th generation.  
For the atmospheric neutrino oscillation, the parameters are expressed by 
\begin{eqnarray}
    \sinatm &=& 4|V_{32}|^2(1-|V_{32}|^2),  \nonumber \\
    \Datm   &=& \mns3-\mns1.  
\end{eqnarray}
Combining Eqs. (\ref{expatm}) and (\ref{expchooz}), 
the magnitude of $V_{31}$ should be small.  
This constraint make it possible to evaluate the parameters of 
the solar neutrino oscillation by  
\begin{eqnarray}
    \sinsol &=& 4|V_{21}|^2(1-|V_{21}|^2),  \nonumber \\
    \Dsol   &=& \mns2-\mns1.  
\end{eqnarray}
These evaluations would be sufficient for our present purpose to 
discuss plausibility of the model at the GUT energy scale.   

\begin{figure}[ht]
\includegraphics[width=8cm,clip]{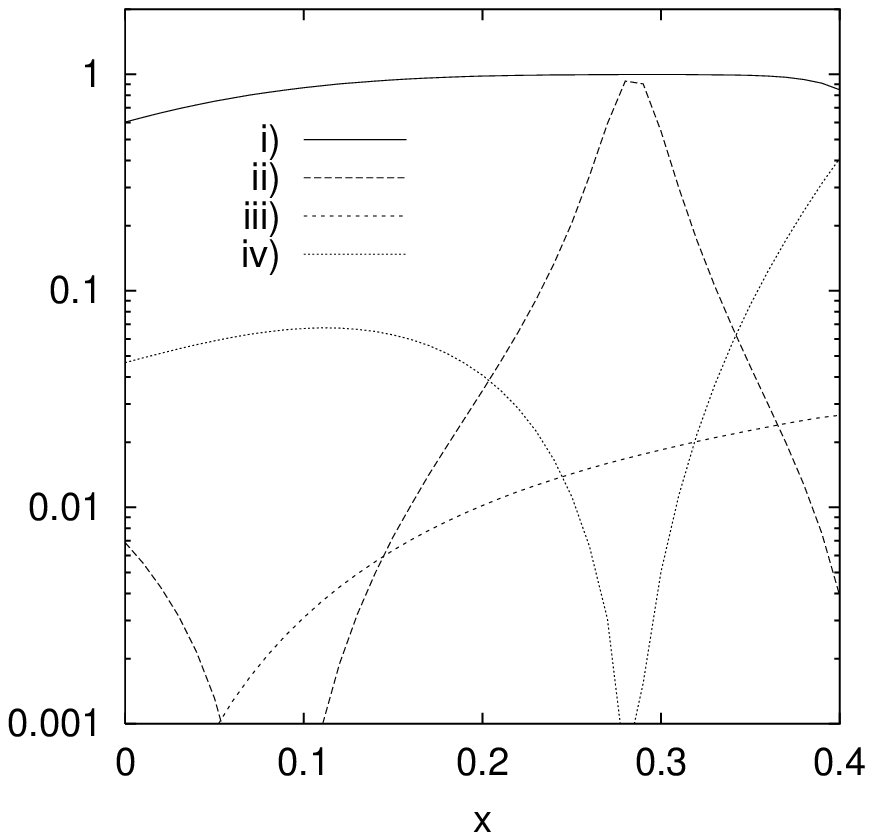}%
 \caption{
The mixing parameters and the ratio of mass-squared differences 
at the electroweak energy scale 
for $C_2^2\epsilon_{23} (\equiv x)$ with 
$C_2^2\epsilon_{12} =C_2^2\epsilon_{13} =0.02$:  
i) $\sinatm$, ii) $\sinsol$, iii) $\sincho$, iv) $\Dsol/\Datm$.  
\label{fig:mixings}}
\end{figure}

\begin{table}
\caption{
The mixing parameters and the ratio of mass-squared differences 
at the electroweak energy scale 
for $C_2^2\epsilon_{12} =0.02$, $C_2^2\epsilon_{13} =0.02$, and 
$C_2^2\epsilon_{23} =0.27$.  
\label{tab:mixings}
}
\begin{ruledtabular}
\begin{tabular}{cccc}
 $\sinatm$ & $\sinsol$ & $\sincho$ & $\Dsol/\Datm$ \\ 
\hline
 1.0  & 0.59 & 0.016 & 3.1$\times 10^{-3}$ \\ 
\end{tabular}
\end{ruledtabular}
\end{table}

     The parameters $a$, $b$, $P_u$, and $P_d$ are constrained from 
Eq. (\ref{electrons}).   
Allowed regions for these parameters are very restricted.  
We take one example within the ranges of real values, 
which is given by 
\begin{eqnarray}
   a &=& -3.95,  \quad b=-10.5,  \nonumber \\
   P_u &=& {\rm diag}(-1,-1,1), \quad  P_d = {\rm diag}(1,1,-1).  
\end{eqnarray} 
This parameter set, together with appropriate values for $\eta_d^D$, 
$\eta_u^D$, $V_{0CKM}$ at the GUT energy scale, leads to the  
quark and charged lepton masses and the CKM matrix at 
the electroweak energy scale listed in Table \ref{tab:massmix}.   
The ratio of the VEVs for $H_1$ and $H_2$ is set for $\tan\beta=30$.  
The obtained results are consistent with the values 
expected at the electroweak energy scale from experiments \cite{fusaoka}.  
The $CP$-violating phase $\delta$ also lies in the range allowed 
by observed $CP$ violation in the $K^0$-$\bar K^0$ and 
$B^0$-$\bar B^0$ systems.  
For a smaller value of $\tan\beta$, the magnitudes of $a$ and $b$ 
become larger.  

     The coefficient matrix $\kappa$ in Eq. (\ref{neutrinos}) is 
now determined by $v_R/C_1^4$ and $C_2^2\epsilon$.  
In Fig. \ref{fig:mixings} the mixing parameters $\sinatm$, 
$\sinsol$, $\sincho$, and the ratio of mass-squared differences 
$\Dsol/\Datm$ at the electroweak energy scale are shown, 
corresponding respectively to curves (i), (ii), (iii), and (iv), 
as functions of $C_2^2\epsilon_{23} (\equiv x)$ within 
the range of real values.  
The other parameters are put at $C_2^2\epsilon_{12}=C_2^2\epsilon_{13}=0.02$.  
The value of $v_R/C_1^4$ determines the scale of $\kappa$ 
and does not affect the four quantities.  
The parameter ranges which give both $\sinsol$ and $\Dsol/\Datm$ 
the values compatible with experiments are not wide.   
In Table \ref{tab:mixings} the resulting values are given for 
$C_2^2\epsilon_{23}=0.27$.  
Putting at $v_R/C_1^4=5.0\times 10^{15}$ GeV, the mass-squared difference 
for the atmospheric neutrino oscillation is given by 
$\Datm=2.7\times 10^{-3}$.  
Under the constraints from the CHOOZ experiment, 
the atmospheric and solar neutrino oscillations can be  
realized simultaneously for certain parameter values.   

     In conclusion, we have presented a model based on SO(10) GUT and 
supersymmetry, in which the quarks and leptons receive masses from the 
Higgs bosons in $\bf 10$, $\bf 120$, and $\overline {\bf 126}$.   
The antisymmetric $\bf 120$ representation is the origin of 
the observed large generation mixings for the leptons.  
The small neutrino masses are traced back to 
large Majorana masses for the right-handed neutrinos generated by 
$\overline {\bf 126}$.  
Theoretical predictions are sensitive to the model parameters.  
All the experimental results can be described 
consistently in some regions of the parameter space.  

\begin{acknowledgments}
     This work is supported in part by the Grant-in-Aid for 
Scientific Research on Priority Areas (No. 14039204) from the 
Ministry of Education, Science and Culture, Japan.   
\end{acknowledgments}


\begin{references}
\bibitem{atm}
     S. Fukuda et al. (Super-Kamiokande Collaboration), \PRL{85}{3999}{2000}.
\bibitem{sol}
   B.T. Cleveland et al., \AJ{496}{505}{1998};  \\
   W. Hampel et al. (GALLEX Collaboration), \PLB{447}{127}{1999};  \\  
   J.N. Abdurashitov et al. (SAGE Collaboration), \PRC{60}{055801}{1999};  \\
   M. Altmann et al. (GNO Collaboration), \PLB{490}{16}{2000};  \\
   S. Fukuda et al. (Super-Kamiokande Collaboration), \PRL{86}{5656}{2001}; \\
   Q.R. Ahmad et al. (SNO Collaboration), \PRL{87}{071301}{2001}.   
\bibitem{dorsner}
    I. Dorsner and S.M. Barr, \NPB{617}{493}{2001}, and references therein.
\bibitem{chooz} 
    M. Apollonio et al., \PLB{466}{415}{1999}.  
\bibitem{babu}
    K.S. Babu and R.N. Mohapatra, \PRL{70}{2845}{1993}; \\
    T. Fukuyama and N. Okada, hep-ph/0205066 (2002).
\bibitem{brahmachari}
     B. Brahmachari and R.N. Mohapatra, \PRD{58}{015001}{1998}.
\bibitem{naculich}
     S.G. Naculich, \PRD{48}{5293}{1993}.
\bibitem{oshimo}
     N. Oshimo, in preparation.
\bibitem{fusaoka}
     See e.g. H. Fusaoka and Y. Koide, \PRD{57}{3986}{1998}.

\end{references}
\end{document}